\documentclass[twocolumn,english,aps,pra,twocolumn,preprintnumbers,amsmath,amssymb]{revtex4-1}
\usepackage[T1]{fontenc}
\usepackage[utf8]{inputenc}
\usepackage{xcolor}
\usepackage{pdfcolmk}
\usepackage{amsmath}
\usepackage{graphicx}
\PassOptionsToPackage{normalem}{ulem}
\usepackage{ulem}

\makeatletter

\providecommand{\tabularnewline}{\\}
\providecolor{lyxadded}{rgb}{0,0,1}
\providecolor{lyxdeleted}{rgb}{1,0,0}

\makeatother

\usepackage{babel}
\begin{document}

\title{Extracting Lyapunov exponents from the echo dynamics of Bose-Einstein
condensates on a lattice}

\author{Andrei E. Tarkhov$^{1}$, Sandro Wimberger$^{2,3,4}$ and Boris V.
Fine$^{1,4}$}

\affiliation{$^{1}$Skolkovo Institute of Science and Technology, Skolkovo Innovation
Center, Novaya~street~100, Skolkovo~143025, Russia}

\affiliation{$^{2}$Dipartimento di Scienze Matematiche, Fisiche e Informatiche,
Università di Parma, Via G.P. Usberti 7/a, I-43124 Parma, Italy}

\affiliation{$^{3}$INFN, Sezione di Milano Bicocca, Gruppo Collegato di Parma,
Italy}

\affiliation{$^{4}$Institute for Theoretical Physics, University of Heidelberg,
Philosophenweg 12, 69120 Heidelberg, Germany}

\date{\today}
\begin{abstract}
We propose theoretically an experimentally realizable method to demonstrate
the Lyapunov instability and to extract the value of the largest Lyapunov
exponent for a chaotic many-particle interacting system. The proposal
focuses specifically on a lattice of coupled Bose-Einstein condensates
in the classical regime describable by the discrete Gross-Pitaevskii
equation. We suggest to use imperfect time-reversal of system's dynamics
known as Loschmidt echo, which can be realized experimentally by reversing
the sign of the Hamiltonian of the system. The routine involves tracking
and then subtracting the noise of virtually any observable quantity
before and after the time-reversal. We support the theoretical analysis
by direct numerical simulations demonstrating that the largest Lyapunov
exponent can indeed be extracted from the Loschmidt echo routine.
We also discuss possible values of experimental parameters required
for implementing this proposal.

\end{abstract}
\maketitle

\section{Introduction}

Historically, statistical physics was established by Boltzmann, Gibbs
and others on the basis of the assumption that the internal dynamics
of a typical interacting many-body system is chaotic. Yet one of the
outstanding issues of the foundations of modern statistical physics
remains to produce experimental evidence that a typical many-particle
system is indeed chaotic. A classical system is called chaotic if
it has at least one positive Lyapunov exponent, which characterizes
exponential sensitivity of phase space trajectories to infinitesimally
small perturbations of initial conditions. The practical challenge
here is that it is impossible: first, to monitor all phase space coordinates
of a many-body system and, second, to prepare initial conditions with
very high accuracy required for extracting Lyapunov exponents. On
top of this, microscopic many-particle systems are not classical,
but quantum, which makes the whole notion of phase space not very
well defined. To make progress on the issue of chaos in statistical
physics, it is reasonable to separate the difficulty of extracting
Lyapunov exponents for classical systems from the difficulty of defining
quantum chaos~\cite{haake2013quantum} as such. In this paper, we
concentrate on the former.

A method of extracting the largest Lyapunov exponent of a many-particle
classical system without using full phase space trajectories was proposed
recently in Ref.~\cite{fine2014absence}. The method is based on
tracking the initial behavior of virtually any observable quantity
in response to imperfect reversal of system's dynamics. This imperfect
reversal is called Loschmidt echo. It can be realized experimentally
by reversing the sign of the Hamiltonian of a system. 

In Ref.~\cite{fine2014absence} the possibility to extract the largest
Lyapunov exponent was demonstrated for a lattice of classical spins,
whereas in the present article we generalize the same analysis to
a system of coupled Bose-Einstein condensates~(BEC) on a lattice
in the regime describable by the classical discrete Gross-Pitaevskii
equation~(DGPE)~\cite{gross1961structure,pitaevskii1961vortex}.
In other words, we consider the classical dynamics of this system,
despite the fact that the system is of quantum origin. The advantage
of coupled Bose-Einstein condensates over classical spins is that
the former were already realized experimentally. In particular, Struck
et al.~\cite{struck2011quantum} have recently performed an experimental
simulation of frustrated classical magnetism using Bose-Einstein condensates
of ultracold atoms. However, Ref.~\cite{struck2011quantum} concentrated
on simulating low-temperature equilibrium properties of the system,
while the present article concentrates on finite-temperature dynamics
and its time-reversal. Time-reversal of DGPE was previously considered
in Refs.~\cite{weiss2012distinguishing,weiss2013effective}, but
not in the context of extracting the largest Lyapunov exponent. An
alternative time-reversal procedure analogous to the sign change of
all particle velocities in classical mechanics was already experimentally
realized for the propagation of a wave-packet of intense light in
a nonlinear crystal, which is describable by the continuous nonlinear
Schrödinger equation, an analog of the continuous Gross-Pitaevskii
equation~\cite{sun2012observation}.

The structure of the present paper is as follows. In Section~\ref{sec:Obtaining-Lyapunov-exponent},
we describe the general idea how one extracts the largest Lyapunov
exponent from Loschmidt echo in a many-particle system. Then, in Section~\ref{sec:Formulation-of-the}
we formally define the problem of Loschmidt echo for interacting BECs
on a lattice. In Section~\ref{sec:Numerical-algorithm}, we provide
some details of the numerical algorithm and describe the methods of
extracting the largest Lyapunov exponent of the system governed by
DGPE in one, two and three dimensions: the direct one and from the
Loschmidt echo. In Section~\ref{sec:Applicability-of-DGPE}, we consider
the limits of applicability of DGPE imposing constraints on experimental
realization. Finally, in Section~\ref{sec:Proposal-of-experiment},
we make a proposal of an experimental setting that could potentially
verify our theoretical results. In particular, we describe the possible
range of system parameters where the approximations we used are valid.

\section{Lyapunov exponent from Loschmidt echo: general idea\label{sec:Obtaining-Lyapunov-exponent}}

In general, a conservative system with $2N$-dimensional phase space
is characterized by a spectrum of $N$ pairs of Lyapunov exponents
of the same absolute value and opposite signs. When two phase space
trajectories $\mathbf{R}_{1}(t)$ and $\mathbf{R}_{2}(t)$ are initially
infinitesimally close to each other, their separation from each other
after sufficiently long time is controlled by the largest positive
Lyapunov exponent $\lambda_{\max}$ of the system. $\lambda_{\max}$
describes the average expansion rate along the direction of the corresponding
eigenvector in tangential space, which typically has fluctuating projections
on all phase space axes. Let us choose one of the axes of the phase
space to correspond to the observable quantity of interest. In such
a case, it is expected that the projection of the difference between
the two separating phase trajectories $\mathbf{R}_{1}(t)$ and $\mathbf{R}_{2}(t)$
on this axis will exhibit erratic behavior, but the envelope of that
behavior will grow exponentially and will be controlled by $\lambda_{\max}$.
If the system is ergodic the value of $\lambda_{\max}$ does not depend
on where the two phase space trajectories start, but the corresponding
eigenvector and the resulting fluctuating projection on the chosen
axis do. It is therefore expected that if one averages over an ensemble
of initial conditions on the same energy shell, then the fluctuating
component of the difference between the trajectories would average
into a constant multiplied by a factor $\exp\left(\lambda_{\max}t\right)$.

As suggested in Ref.~\cite{fine2014absence}, the above considerations
can be converted into the following scheme of extracting $\lambda_{\max}$.
Let us consider equilibrium noise of observable $X$ as a function
of time $t$ for a system governed by Hamiltonian ${\cal H}$. Next,
we record this noise during time-interval from $0$ to $\tau$ and
at time $\tau$ reverse the sign of the Hamiltonian with a slight
perturbation of the system at the moment of Hamiltonian reversal.
If the perturbation is infinitesimally small, the quantity $X(\tau+\Delta t)$
will be tracking the quantity $X(\tau-\Delta t)$, while gradually
departing from it as the echo time $\Delta t$ increases. After sufficiently
long time, $\left|X(\tau+\Delta t)-X(\tau-\Delta t)\right|$ should
be modulated by $\exp\left(\lambda_{\max}\Delta t\right)$. The preceding
consideration then suggests that $\lambda_{\max}$ can be extracted
from the following average over the initial conditions

\begin{equation}
\lambda_{\max}=\frac{1}{\Delta t}\left\langle \log\left|X(\tau+\Delta t)-X(\tau-\Delta t)\right|\right\rangle ,\label{eq:LE_definition}
\end{equation}
where $\tau$ should be larger than $\Delta t$~\footnote{Ref.~\cite{fine2014absence} used the logarithm of ensemble average
to obtain the largest Lyapunov exponent, as opposed to the average
of the logarithm defined by Eq.~(\ref{eq:LE_definition}). Both procedures
are equivalent in the thermodynamic limit $N\to\infty$. However,
for not too large systems, the logarithm of the average leads to a
systematic correction in the value of $\lambda_{\max}$.\label{fn:Ref.-used-the}}.

\begin{figure}[h]
\includegraphics[width=1\columnwidth]{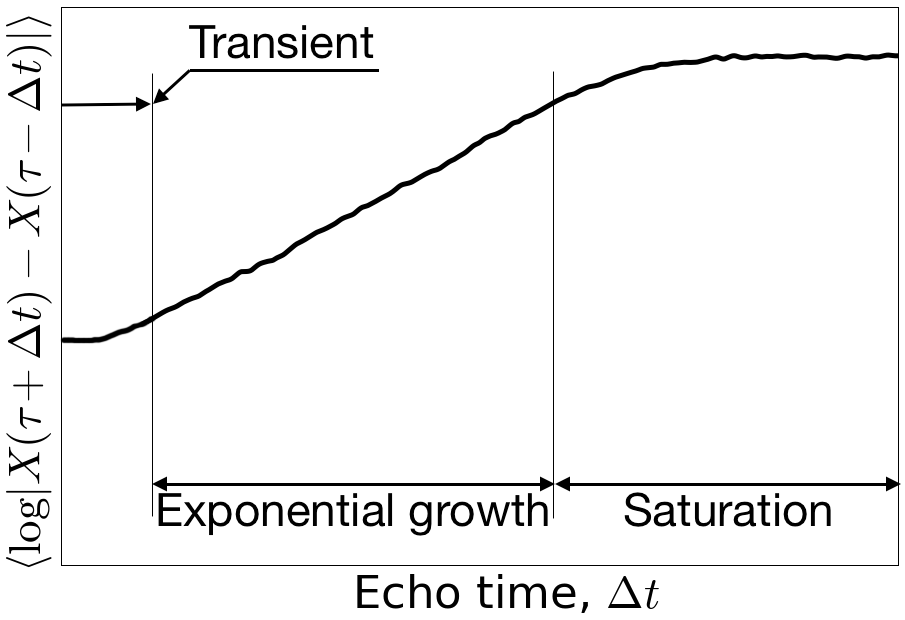}

\caption{Sketch of a typical Loschmidt echo response $\left\langle \log\left|X(\tau+\Delta t)-X(\tau-\Delta t)\right|\right\rangle $~(thick
black line). Three characteristic regimes described in the text are
indicated: transient, exponential growth and saturation. \label{fig:sketch_for_LE}}
\end{figure}

Typical behavior of $\left\langle \log\left|X(\tau+\Delta t)-X(\tau-\Delta t)\right|\right\rangle $
as a function of $\Delta t$ for almost any reasonable quantity $X$
is qualitatively depicted in Fig.~\ref{fig:sketch_for_LE}. It starts
growing from a tiny value at $\Delta t=0$ and then evolves through
a transient regime, where all Lyapunov exponents contribute to the
growth, and the largest one is not dominant yet. After that, it enters
the exponential growth regime, where the largest Lyapunov exponent
controls the growth. For any finite initial difference between the
two departing phase space trajectories, the exponential growth regime
is eventually followed by the saturation regime, where $\left|X(\tau+\Delta t)-X(\tau-\Delta t)\right|$
is no longer small enough to be describable by linearized dynamics.
This means that, experimentally or numerically, the perturbation of
a perfect time-reversal should be small enough, so that the time $\Delta t$
for which $\left|X(\tau+\Delta t)-X(\tau-\Delta t)\right|$ remains
small is sufficiently long to extract $\lambda_{\max}$.

As follows from the above analysis, the method does not use any specific
properties of quantity $X$, thus it can be either scalar or vector.
If one chooses a $K-$dimensional vector observable $\mathbf{X}=\{X_{i}\}$,
then the perturbation of interest $\left|\mathbf{X}(\tau+\Delta t)-\mathbf{X}(\tau-\Delta t)\right|$
can be redefined as $\sqrt{\sum_{i=1}^{K}\left(X_{i}(\tau+\Delta t)-X_{i}(\tau-\Delta t)\right)^{2}}$.

We finally remark, that, as demonstrated in Ref.~\cite{fine2014absence},
the qualitative picture of the three regimes, that are sketched in
Fig.~\ref{fig:sketch_for_LE}, remains valid also when the perturbation
making the time-reversal imperfect comes not only from a small shaking
of the system at time $\tau$, but also from an imperfect reversal
of system's Hamiltonian.

\section{Formulation of the problem\label{sec:Formulation-of-the}}

In this work we consider Bose-Einstein condensates on a lattice of
$N$ sites describable by DGPE

\begin{equation}
i\frac{d\psi_{j}}{dt}=-J\sum_{k}^{N\!N(j)}\psi_{k}+\beta\left|\psi_{j}\right|^{2}\psi_{j},\label{eq:equations-motion-DGPE}
\end{equation}
where $\psi_{j}$ is the complex order-parameter, describing the condensate
at site $j=1\ldots N$, $J$ and $\beta$ are two parameters, controlling
hopping and nonlinear on-site interactions, respectively. The summation
over $k$ extends over the nearest-neighbors $NN(j)$ of site $j$.
As shown in Section~\ref{sec:Proposal-of-experiment}, DGPE is derivable
from the Bose-Hubbard model in the limit of large occupation numbers.

DGPE generates conservative dynamics corresponding to the Hamiltonian

\begin{equation}
{\cal H}=-J\sum_{\left\langle i,j\right\rangle }\psi_{i}^{*}\psi_{j}+\frac{\beta}{2}\sum_{i}\left|\psi_{i}\right|^{4}.\label{eq:Gross-Pitaevskii-equation-NN}
\end{equation}
This dynamics has two integrals of motion: the total energy $E_{total}$~(the
r.h.s. of Eq.~(\ref{eq:Gross-Pitaevskii-equation-NN})) and the total
number of particles $N_{p}=\sum_{i}\left|\psi_{i}\right|^{2}$. 

For all our calculations, we have chosen $J=1$, $\beta=0.01$ and
the initial conditions $\left|\psi_{i}(0)\right|^{2}=100$ with almost
random phases, fixed such that the energy per site is equal to $100$
by the procedure described in Section~\ref{sec:Numerical-algorithm}.
With the above choice, the energy is nearly equally distributed between
different sites and between the hopping and the interaction terms
in Eq.~(\ref{eq:Gross-Pitaevskii-equation-NN}). This allows the
system to stay in the ergodic regime not influenced by solitonic and
breather-like solutions. (The experience with classical spin lattices~\cite{de2012largest,de2013lyapunov}
indicates that many-body classical systems are generically ergodic
and chaotic at energies corresponding to sufficiently high temperatures.)

\begin{figure}[t]
\begin{raggedright}
(a)
\par\end{raggedright}
\begin{raggedright}
\includegraphics[width=1\columnwidth]{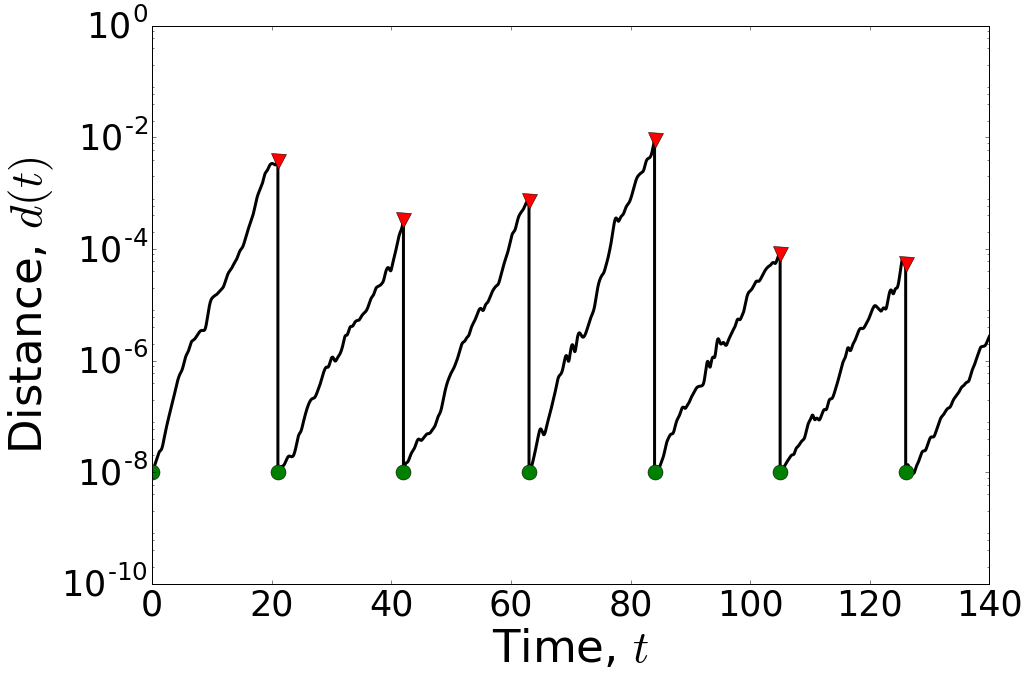}
\par\end{raggedright}
\begin{raggedright}
(b)
\par\end{raggedright}
\raggedright{}\includegraphics[width=1\columnwidth]{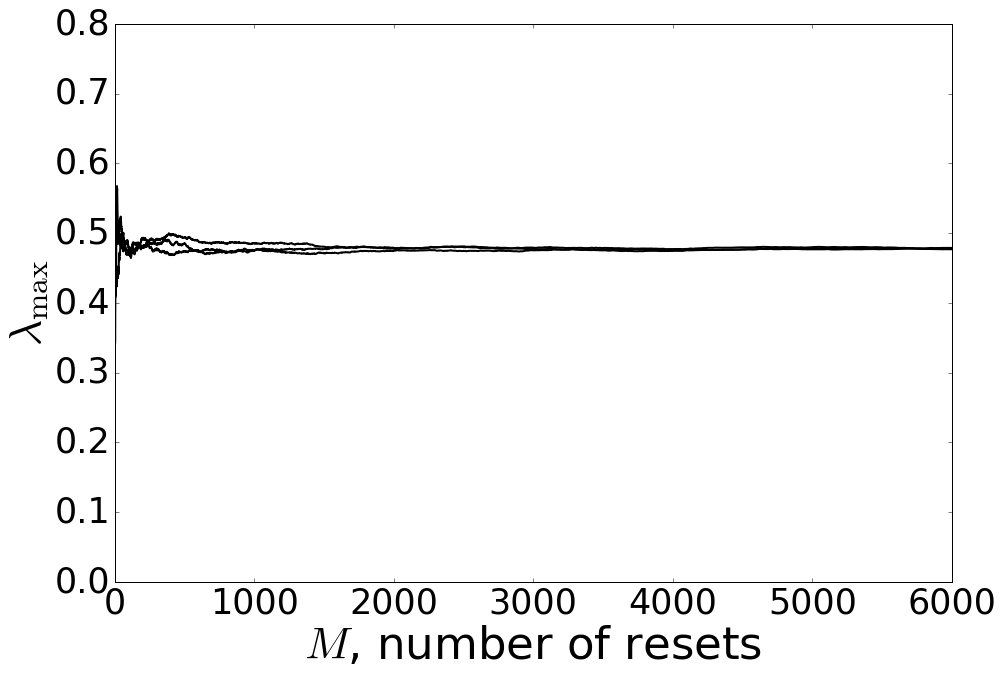}\caption{(Color online) Illustrations of the numerical routine for computing
$\lambda_{\max}$. (a) Black line: distance $d(t)=\left\Vert \mathbf{R}_{1}(t)-\mathbf{R}_{2}(t)\right\Vert _{2}$
between two phase space trajectories $\mathbf{R}_{1}(t)$ and $\mathbf{R}_{2}(t)$
used for computing $\lambda_{\max}$ for DGPE on a one-dimensional
lattice with $N=10$ sites. Time is divided into intervals of duration
$T_{0}\approx20$, each starts at the reset time $t_{m}$~(green
dots), for which $d(t_{m})=d_{0}=10^{-8}$, and finishes at time $t_{m}+T_{0}$~(red
triangles). According to Eq.~(\ref{eq:LLE_calculation_numerical}),
the contribution to $\lambda_{\max}$ from each such an interval (local
stretching rate) is $\frac{1}{T_{0}}\log\left|\frac{d(t_{m})}{d_{0}}\right|$.
(b) Ergodicity test: lines represent $\lambda_{\max}$ obtained from
Eq.~(\ref{eq:LLE_calculation_numerical}) as a function of the number
of resets $M$. Each line is obtained for different randomly-chosen
initial conditions for $\mathbf{R}_{1}(t)$ on the same shell of constant
$E_{total}$ and $N_{p}$. Convergence to a single value of $\lambda_{\max}$
with time indicates that the system is ergodic. \label{fig:Ergodicity}\label{fig:Local_stretching_rates}}
\end{figure}

We mark all the variables corresponding to the time interval preceding
the time-reversal at time $\tau$ with a subscript ``$-$'' and
succeeding the time-reversal with a subscript ``$+$''.

Loschmidt echo is implemented as follows. The time evolution of the
system during time interval~$[0,\tau]$ is governed by the Hamiltonian
${\cal H}_{-}$~(\ref{eq:Gross-Pitaevskii-equation-NN}) and, after
time $\tau$, by the sign-reversed Hamiltonian $\mathcal{H}_{+}=-{\cal H}_{-}$,
i.e. we change the sign of the Hamiltonian parameters at time $\tau$:
$J_{+}=-J_{-}$, $\beta_{+}=-\beta_{-}$. How to realize such a time-reversal
experimentally will be discussed in Section~\ref{subsec:Time-reversal-of-dynamics}.
At the moment of time reversal, we also introduce a tiny perturbation
to the state vector: $\psi_{i}(\tau+0)=\psi_{i}(\tau-0)+\delta\psi_{i}$,
where $\left\{ \delta\psi_{i}\right\} $ is a random vector, subject
to the constraint $\sqrt{\sum_{i}\left|\delta\psi_{i}\right|^{2}}=10^{-8}$.

We have chosen a set of on-site occupations $n_{i}\equiv\left|\psi_{i}^{2}\right|$
as the quantity of interest $\mathbf{X}(t)\equiv\{n_{1},n_{2},\ldots,n_{N}\}$.
Thus, we characterize Loschmidt echo by the function $G(\Delta t)\equiv\left\langle \log\left|\mathbf{X}(\tau+\Delta t)-\mathbf{X}(\tau-\Delta t)\right|\right\rangle $,
which for the for the chosen quantity of interest can be written as

\begin{equation}
G(\Delta t)=\left\langle \log\sqrt{\sum_{i=1}^{N}\left[\Delta n_{i}(\Delta t)\right]^{2}}\right\rangle ,\label{eq:Loschimdt_echo_for_populations}
\end{equation}
where $\Delta n_{i}(\Delta t)\equiv n_{i}(\tau+\Delta t)-n_{i}(\tau-\Delta t)$,
and $\left\langle \ldots\right\rangle $ denotes ensemble averaging
over initial conditions. As explained in Section~\ref{sec:Obtaining-Lyapunov-exponent},
the regime of the exponential growth of perturbation is expected to
be characterized by the asymptotic relation 

\begin{equation}
G(\Delta t)\cong\lambda_{\max}\Delta t,\label{eq:asymptotic_G_delta_t}
\end{equation}
from which the value of the largest Lyapunov exponent can be extracted.
In the following sections, we demonstrate the validity of the above
proposition by, first, directly calculating $\lambda_{\max}$ according
to the algorithm of Ref.~\cite{sprott2003chaos}, and then comparing
it with the value extracted from Eq.~(\ref{eq:asymptotic_G_delta_t})
on the basis of direct simulations of Loschmidt echoes.

We will do this for a one-dimensional lattice with $10$ sites, a
two-dimensional square lattice of size $10\times10$ and a three-dimensional
cubic lattice of size $4\times4\times4$ with nearest-neighbor interactions
and periodic boundary conditions.

\section{Numerical algorithm\label{sec:Numerical-algorithm}}

To simulate the solutions of DGPE, we employ a Runge-Kutta 4th order
algorithm with discretization step $\delta t$=0.001. This limits
the algorithmic error to $O(\delta t^{4})$ or roughly $10^{-12}$,
whereas by using the quadrupole-precision numbers we fix the machine
precision to be roughly $10^{-33}$. 

The value of $\lambda_{\max}$ in general depends on the two conserved
quantities of the system $E_{total}$ and $N_{p}$.

We generate an ensemble of initial conditions corresponding to $E_{total}=100N$
and $N_{p}=100N$, where $N$ is the number of lattice sites. We do
this by choosing initially all $\left|\psi_{i}\right|=10$, with random
phases. Then, we minimize $\left(E_{total}-100N\right)^{2}+\left(N_{p}-100N\right)^{2}$
by the steepest descent optimization procedure.

As mentioned in Section~\ref{sec:Formulation-of-the}, we introduce
a small perturbation at the moment of time-reversal by adding a random
perturbation $\left\{ \delta\psi_{i}\right\} $ to the state vector
$\left\{ \psi_{i}\right\} $. The length of the perturbation vector
is $10^{-8}$. This procedure slightly changes $E_{total}$ and $N_{p}$,
but the resulting difference in the value of the largest Lyapunov
exponent is several orders of magnitude smaller than the chosen precision
of $3$ significant digits. Therefore, we can neglect it.

For further details one can refer to the source code published in
a GitHub repository~\footnote{The code used for the analysis in the present paper is provided in
a GitHub repository at https://github.com/TarkhovAndrei/DGPE\label{fn:github}}.

\subsection{Lyapunov exponent calculation\label{sec:Lyapunov-exponent-calculation}}

The definition of the largest Lyapunov exponent reads

\begin{equation}
\lambda_{\max}\equiv\frac{1}{t}\lim\left(\log\left|\frac{d(t)}{d(0)}\right|\right)_{t\to\infty,d(0)\to0},\label{eq:LLE_definition}
\end{equation}

where $d(t)=\left\Vert \mathbf{R}_{1}(t)-\mathbf{R}_{2}(t)\right\Vert _{2}$
is the distance between two phase space trajectories, which are infinitesimally
close to each other at $t=0$.

This definition is not practical for numerical simulation because
it in general requires unachievable computational precision. Instead,
we perform the direct calculation of the largest Lyapunov exponent
$\lambda_{\max}$ following the standard numerical algorithm, see
e.g. Ref.~\cite{wimberger2014nonlinear}.

This algorithm tracks two trajectories: the reference trajectory $\mathbf{R}_{1}(t)$
and the slightly perturbed trajectory $\mathbf{R}_{2}(t)=\mathbf{R}_{1}(t)+\mathbf{\delta R}(t)$.
The algorithm starts with $\left|\mathbf{\delta R}(0)\right|=d_{0}$
and then lets $\mathbf{\delta R}(t)$ grow during time interval $T_{0}$,
then it shifts $\mathbf{R}_{2}(t)$ closer to $\mathbf{R}_{1}(t)$
by resetting the length of $\mathbf{\delta R}$ back to $d_{0}$.
This procedure is repeated as many times as necessary, until the following
quantity converges:

\begin{equation}
\lambda_{max}=\frac{1}{MT_{0}}\sum_{m}^{M}\log\left|\frac{d(t_{m})}{d_{0}}\right|,\label{eq:LLE_calculation_numerical}
\end{equation}
where $M$ is the number of resets, $m$ is the reset index, $t_{m}$
is the time just before the $m$-th reset. The time evolution of the
distance $d(t)$ in the course of such simulation is presented in
Fig.~\ref{fig:Local_stretching_rates}(a).

In all our simulations we test the ergodicity of system's dynamics
numerically by checking that the values of $\lambda_{\max}$ obtained
for several randomly chosen initial conditions on a shell with the
given values of $E_{total}$ and $N_{p}$ are the same. In all cases
reported below, this ergodicity test was positive. One such a test
is illustrated in Fig.~\ref{fig:Ergodicity}(b).

\subsection{Loschmidt echo simulations\label{sec:Loschmidt-echo-calculation}}

\begin{figure}[h]
\includegraphics[width=1\columnwidth]{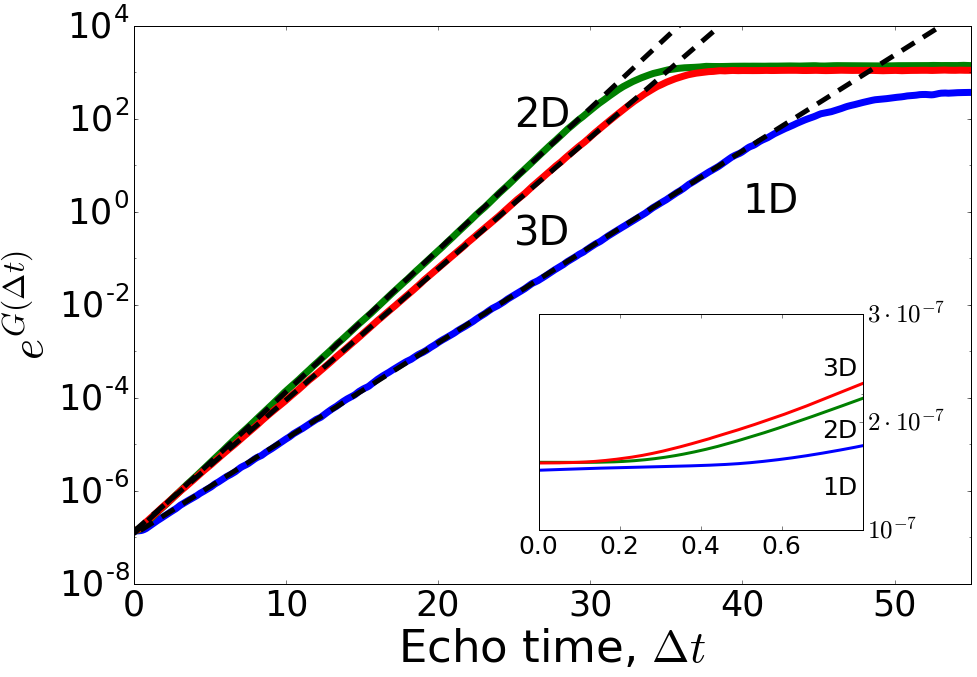}\caption{(Color online) Loschmidt echo response $\exp\left(G(\Delta t)\right)$
obtained from Eq.~(\ref{eq:Loschimdt_echo_for_populations}) for
a one-dimensional chain of $10$ sites~(1D, blue line), a two-dimensional
$10\times10$ square lattice~(2D, green line) and a three-dimensional
$4\times4\times4$ cubic lattice~(3D, red line). The inset shows
the behavior of $\exp\left(G(\Delta t)\right)$ in the transient regime
at small echo times, where all Lyapunov exponents contribute to the
growth. The transient regime takes longer time for lower dimensions.
In Table~\ref{tab:Comparison-of-the}, the values of $\lambda_{\max}$
obtained by fitting the exponential growth regime are compared to
those obtained from the direct calculation described in Section~\ref{sec:Lyapunov-exponent-calculation}.
\label{fig:Loschmidt_echo_for_dN_vs_LLE}}
\end{figure}

We have computed the Loschmidt echo response function $G(\Delta t)$
given by Eq.~(\ref{eq:Loschimdt_echo_for_populations}) for one-,
two- and three-dimensional lattice geometries with the parameters
defined in Section~\ref{sec:Formulation-of-the}. The results of
these simulations are presented in Fig.~\ref{fig:Loschmidt_echo_for_dN_vs_LLE}.

As clearly seen in Fig.~\ref{fig:Loschmidt_echo_for_dN_vs_LLE},
the expected exponential growth regime of $G(\Delta t)$ is present
in all three cases. The values of $\lambda_{\max}$ characterizing
this regime are summarized in Table.~\ref{tab:Comparison-of-the},
where they are also compared with the values of $\lambda_{\max}$
obtained from the direct calculation described in Section~\ref{sec:Lyapunov-exponent-calculation}.
The agreement between the two sets of values is within the numerical
accuracy of the calculations. Similar agreement was demonstrated previously
in Ref.~\cite{fine2014absence} for classical spins. We finally note
here that the fact that the largest Lyapunov exponent for the 3D cubic
lattice is slightly smaller than that for the 2D square lattice is
presumably a finite size effect related to the small size of the 3D
lattice.

\begin{table}[h]
\begin{tabular}{|c|c|c|}
\hline 
 & $\lambda_{\max}$ from direct calculation & $\lambda_{\max}$ from Loschmidt echo\tabularnewline
\hline 
\hline 
1D & $0.481\pm0.002$ & $0.475\pm0.004$\tabularnewline
\hline 
2D & $0.703\pm0.003$ & $0.702\pm0.004$\tabularnewline
\hline 
3D & $0.648\pm0.002$ & $0.650\pm0.003$\tabularnewline
\hline 
\end{tabular}\caption{Comparison of the largest Lyapunov exponents $\lambda_{\max}$ obtained
from the direct calculation with those extracted from Loschmidt echoes
shown in Fig.~\ref{fig:Loschmidt_echo_for_dN_vs_LLE} for one-, two-
and three-dimensional lattices.\label{tab:Comparison-of-the}}
\end{table}

\section{Applicability of DGPE as a constraint on experimental implementation\label{sec:Applicability-of-DGPE}}

Throughout the paper we used DGPE to model the dynamics of Bose-Einstein
condensates on a lattice. In order to observe experimentally the regime
of exponential growth $G(\Delta t)$ and to extract from this regime
the value of $\lambda_{\max}$, the measured system should be such
that DGPE approximates its dynamics with a very high accuracy. The
question then arises whether such an accuracy is feasible for realistic
experimental settings. To address this question, let us recall that
DGPE is normally justified for the lattices of Bose-Einstein condensates
by describing it at a level of a more fundamental Bose-Hubbard model.
Therefore, we have to define the experimental regime, where both conditions
would be satisfied simultaneously: the Bose-Hubbard model would be
applicable and the classical mean-field approximation to it would
be sufficiently accurate.

The Bose-Hubbard model is defined by the Hamiltonian

\begin{equation}
\hat{{\cal H}}_{BH}=-J\sum_{\left\langle i,j\right\rangle }\hat{a}_{i}^{+}\hat{a}_{j}+\frac{\beta}{2}\sum_{i}\hat{n}_{i}\hat{n}_{i},\label{eq:Bose-Hubbard-model}
\end{equation}
where $\hat{a}_{i}^{+}$ and $\hat{a}_{i}$ are the quantum creation
and annihilation operators for site $i$ respectively, $\hat{n}_{i}\equiv\hat{a}_{i}^{+}\hat{a}_{i}$
is the operator for the occupation at site $i$, $J$ is the hopping
parameter, $\beta$ is the on-site interaction parameter, and the
notation $\left\langle i,j\right\rangle $ implies nearest-neighbor
sites. When the number of bosons in each potential well is large,
one can approximate the Bose-Hubbard Hamiltonian~(\ref{eq:Bose-Hubbard-model})
with the DGPE Hamiltonian~(\ref{eq:Gross-Pitaevskii-equation-NN})
by making the following substitution: $\hat{a}_{i}=\psi_{i}$, $\hat{a}_{i}^{+}=\psi_{i}^{*}$,
$\hat{n}_{i}=n_{i}=\left|\psi_{i}\right|^{2}$.

For the single-orbital Bose-Hubbard model to be valid, the hopping
term $J$ must be relatively small, so that the lattice potential
is deep enough and, as a result, the gap $\Delta_{0}$ between the
lowest and the second lowest bands is sufficiently large~\cite{morsch2006dynamics,bloch2008many}.
In addition, in order for a Lyapunov instability to be observable,
not only the order-parameters $\psi_{i}$ but also small deviations
$\delta\psi_{i}$ should be well defined in the mean-field approximation,
which implies sufficiently large values of $n_{i}$. The implementation
of our proposal then requires the following conditions to be satisfied:
(i) $J\ll\Delta_{0}$ — the condition for not involving the second
band, (ii) $\beta n_{j}\lesssim J$ — the condition preventing the
system from exhibiting self-trapping~\cite{milburn1997quantum,smerzi1997quantum,raghavan1999coherent,leggett2001bose,albiez2005direct,khomeriki2007driven},
(iii) ideally, the number of particles per well $n_{i}$ should be
of the order of $500$ or larger~\cite{morsch2006dynamics,bloch2008many,vardi2001bose}.
We note here that (ii) together with (iii) imply that the condition
for the validity of the mean-field approximation in the Bose-Hubbard
model, $\beta/J\ll1$, is automatically fulfilled. It should be possible
to satisfy all the above conditions with an optical lattice having
potential depth of the order of $5\div10$ recoil energies and not
too strong interactions between atoms~\cite{holthaus2000bloch}.
We also note that the numerical experience with large quantum spins~\cite{elsayed2015sensitivity}
indicates that even $n_{i}\sim15$ might be already sufficient to
extract the largest Lyapunov exponent.

\section{Experimental Proposal \label{sec:Proposal-of-experiment}}

An experiment implementing our proposal should satisfy the following
requirements: (i)~high accuracy of the measurements of the number
of particles $n_{i}$ for individual sites leading to the high accuracy
of $G(\Delta t)$ extracted from these measurements, (ii)~high accuracy
of the experimental realization of the time-reversed Hamiltonian and
(iii)~high accuracy of the DGPE approximation for the given experimental
setting. The relative accuracy in each case should be at least $10^{-2}$
and preferably better. Let us now consider the above requirements
one by one.

\subsection{Measurement of the quantity of interest}

In order to extract $G(\Delta t)$ from experiment, the initial and
the final values of $n_{i}$ should be measured with high accuracy.
In principle, there exist techniques, such as the absorption imaging~\cite{muessel2013optimized}
or the resonant fluorescence detection~\cite{hume2013accurate} that
allow one to achieve the required accuracy. In particular, the current
state-of-the-art record for the resonant fluorescence detection~\cite{hume2013accurate}
is to measure the number of atoms of the order of one thousand with
accuracy better than one percent. However, our proposal implies an
additional requirement, namely, that the initial measurement should
not significantly perturb $n_{i}$, so that the measured values represent
the initial conditions for the actual experimental run. This implies
that destructive techniques, such as absorption imaging, would not
be suitable for the initial measurement, because they would destroy
the condensate. Therefore, it is preferable that at least the initial
measurement is performed by a non-destructive technique, such as,
e.g., dispersive~(off-resonance) imaging~\cite{andrews1996direct,miesner1998bosonic}
or the techniques used in Refs.~\cite{wilson2015situ,figl2006demonstration}.
The alternative approach would be to controllably prepare the initial
state with an accurate \textit{a priori} knowledge of the initial
number of particles on each site. The final measurement can then be
done by either destructive or non-destructive imaging technique.

\subsection{Initial and final conditions}

We propose to create the optical lattice initially with sufficiently
high potential barriers between adjacent sites, which would suppress
hopping between them while the initial occupations are measured. Then,
the barriers should be lowered to the heights corresponding to the
desired value of the hopping parameter $J$. The barriers should be
lowered sufficiently fast, so that the initial occupations of individual
wells remain the same. At the same time, after the barriers are lowered
the initial phases of individual order-parameters $\psi_{i}$ are
expected to be random. Thereby an ensemble of random initial conditions
is to be implemented. After this, both the direct and the reversed
time evolution should last for a time $\tau$ each. Then, the barriers
should be raised again, so that the final occupations of individual
wells can be measured slowly and accurately.

\subsection{Time-reversal of dynamics\label{subsec:Time-reversal-of-dynamics}}

In order to reverse the sign of the Hamiltonian $\mathcal{H}$~(\ref{eq:Gross-Pitaevskii-equation-NN})
at time $\tau$, one can change the sign of the hopping parameter
$J$ and the interaction parameter $\beta$.

The sign-reversal of $J$ can be implemented using fast periodic shaking
of the optical lattice. As shown in Refs.~\cite{eckardt2005superfluid,lignier2007dynamical,struck2011quantum},
the effective hopping parameter $J$ depends on the periodic forcing
amplitude $F$ and the modulation frequency $\omega$ as follows:

\begin{equation}
J(F,\omega)={\cal J}_{0}\left(\frac{d\left|F\right|}{\hbar\omega}\right)\tilde{J},\label{eq:hopping_reversal}
\end{equation}
where ${\cal J}_{0}$ is the zeroth order Bessel function, $\tilde{J}$
is the bare hopping parameter and $d$ is the lattice spacing. Since
${\cal J}_{0}$ is a sign-alternating function, one can find pairs
of parameters $F_{1}$, $\omega_{1}$ and $F_{2}$, $\omega_{2}$,
such that $J(F_{2},\omega_{2})=-J(F_{1},\omega_{1})$. Such a time-reversal
can be implemented on the timescale of the order of the modulation
frequency $\omega$, which is several kHz~\cite{sias2007resonantly,zenesini2009time,struck2011quantum}.

The sign-reversal of the interaction parameter $\beta$ can be implemented
with the help of Feshbach resonances~\cite{feshbach1958unified,inouye1998observation}.
This parameter is proportional to the atomic $s$-wave scattering
length $a_{sc}$, whose value and sign can be controlled by the value
of external magnetic field $B$. Cesium or rubidium 85 could be good
candidates for this kind of experiment, due to their broad Feshbach
resonances~\cite{roberts1998resonant,claussen2003very,gustavsson2008control}.
In this case, the on-site interaction can be reversed on a timescale
of fractions of ms.

According to the above proposal, the time-reversal of the effective
Hamiltonian $\mathcal{H}$ can be implemented within a fraction of
ms, whereas the system dynamics controlled by the values of $J$ and
$\beta$ can be at least one order of magnitude slower.

Bose-Einstein condensates with attractive interaction~(which will
be required either for the forward or the backward time-evolution)
are in general unstable to collapse. However, if they are constrained
to a finite volume, the collapse happens only for numbers of atoms
above a certain critical value, which for realistic optical lattice
parameters can be above 1000 per lattice site~\cite{sackett1999measurements,gerton2000direct,donley2001dynamics,khaykovich2002formation,strecker2002formation,eigen2016observation}.
As mentioned earlier, the implementation of our proposal requires
about 500 atoms per lattice site.

Another useful possibility that potentially improves the flexibility
of experimental implementation is to achieve the time-reversal not
by realizing the strict condition $\mathcal{H}_{+}=-\mathcal{H}_{-}$,
but, instead, borrowing the idea from the magic echo of nuclear magnetic
resonance~\cite{rhim1970violation,slichter2013principles}, to change
the sign of $J$ and $\beta$ in such a way that the Hamiltonian before
the time-reversal $J_{-},\beta_{-}$ are related to the parameters
after the time-reversal $J_{+},\beta_{+}$ as follows: $J_{+}=-CJ_{-}$,
$\beta_{+}=-C\beta_{-}$, where $C$ is some positive constant. In
such a case, $\mathcal{H}_{+}=-C\mathcal{H}_{-}$, so that the time-reversal
routine would consist of the direct time-evolution taking time $\tau$
and the reversed time-evolution taking time $\tau/C$.

\subsection{Lattice geometry}

Experimentally realized optical lattices are, normally, not quite
translationally invariant because of the presence of physical borders.
This, in particular, leads to an effective position-dependent on-site
potential and/or position-dependent hopping, whose values near the
borders of the lattice are different from those in the bulk. In such
a case, the time-reversal of the full Hamiltonian requires reversing
the sign of the above position-dependent terms, which, in turn, poses
an additional experimental complication. It is, therefore, preferable
for implementing our proposal to use an optical lattice that actually
has periodic boundary conditions, which, for all practical purposes,
leaves us with a ring-shaped one-dimensional lattice. Such a lattice
can be realized, for example, on the basis of an interference pattern
of two Laguerre-Gauss modes with different orbital indices~\cite{franke2007optical,amico2014superfluid}.

\section{Conclusions}

We proposed a method to extract the largest Lyapunov exponent for
a lattice of Bose-Einstein condensates on the basis of a Loschmidt
echo routine. We have validated this method by numerical simulations
and discussed its possible experimental implementation with ultracold
bosonic atoms in optical lattices. A successful realization of this
proposal may produce a long-sought direct experimental evidence that
the dynamics of a typical many-particle system is chaotic. This, in
turn, would put the theory of dynamic thermalization on a firmer foundation.
\begin{acknowledgments}
This work was supported by grant of Russian Science Foundation (project
number 17-12-01587).
\end{acknowledgments}

\appendix

\bibliographystyle{apsrev4-1}
\bibliography{references}

\end{document}